\documentclass[twocolumn,final]{autart}    

\usepackage{graphicx}          
  \usepackage{amssymb}
 \usepackage{amsmath}
\usepackage{mathptmx}
\usepackage{times}

\usepackage[colorlinks,
            linkcolor=red,
            anchorcolor=blue,
            citecolor=blue
            ]{hyperref}
\usepackage{cite}

\newtheorem{remark}{Remark}
\newtheorem{theorem}{Theorem}

\newtheorem{definition}{Definition}

\newtheorem{assumption}{Assumption}

\begin{document}

\begin{frontmatter}

\title{Quantum Optical Realization  of  Classical Linear Stochastic Systems\thanksref{footnoteinfo}} 

\thanks[footnoteinfo]{The work was supported by AFOSR Grants FA2386-09-1-4089 and FA2386-12-1-4075, and the
    Australian Research Council.\\ {\normalsize$^{*}$}~~Corresponding author. Tel: +61 261258826.}
\author[rvt]{Shi Wang}$^{,*}$,
\ead{shi.wang@anu.edu.au}
\author[ccc]{Hendra I. Nurdin},
\ead{h.nurdin@unsw.edu.au}
\author[cdc]{Guofeng Zhang },
\ead{Guofeng.Zhang@polyu.edu.hk}
\author[focal]{Matthew R. James}
\ead{Matthew.James@anu.edu.au}

\address[rvt]{Research School of
    Engineering, Australian
    National University, Canberra, ACT 0200,
         Australia.}
\address[ccc]{School of
    Electrical Engineering and Telecommunications, University of New South Wales, Sydney,
N.S.W. 2052
Australia.}
\address[cdc]{Department of Applied Mathematics
The Hong Kong Polytechnic University, Hung Hom, Kowloon,
HKSAR, China}
\address[focal]{Centre for Quantum
Computation and Communication Technology, Research  School of
    Engineering, Australian
    National University, Canberra, ACT 0200,
    Australia.}
\date{}

\begin{keyword}                           
Classical linear stochastic system; Quantum optics; Measurement feedback control; Quantum optical realization.               
\end{keyword}                             

\begin{abstract}                          
The purpose of this paper is to show how a class of classical linear
stochastic systems can be physically implemented using quantum
optical components.
Quantum optical systems typically have much
higher bandwidth than electronic devices, meaning faster response
and processing times, and hence has the potential for providing
better performance than classical systems.
A procedure is provided
for constructing the quantum optical realization. The paper also
describes the use of the quantum optical realization in a
measurement feedback loop. Some examples are given to illustrate the
application of the main results.
\end{abstract}

\end{frontmatter}

\section{Introduction and Motivation}
\label{sec:intro}

\begin{figure}[htbp]
\centering
\includegraphics[scale=0.41]{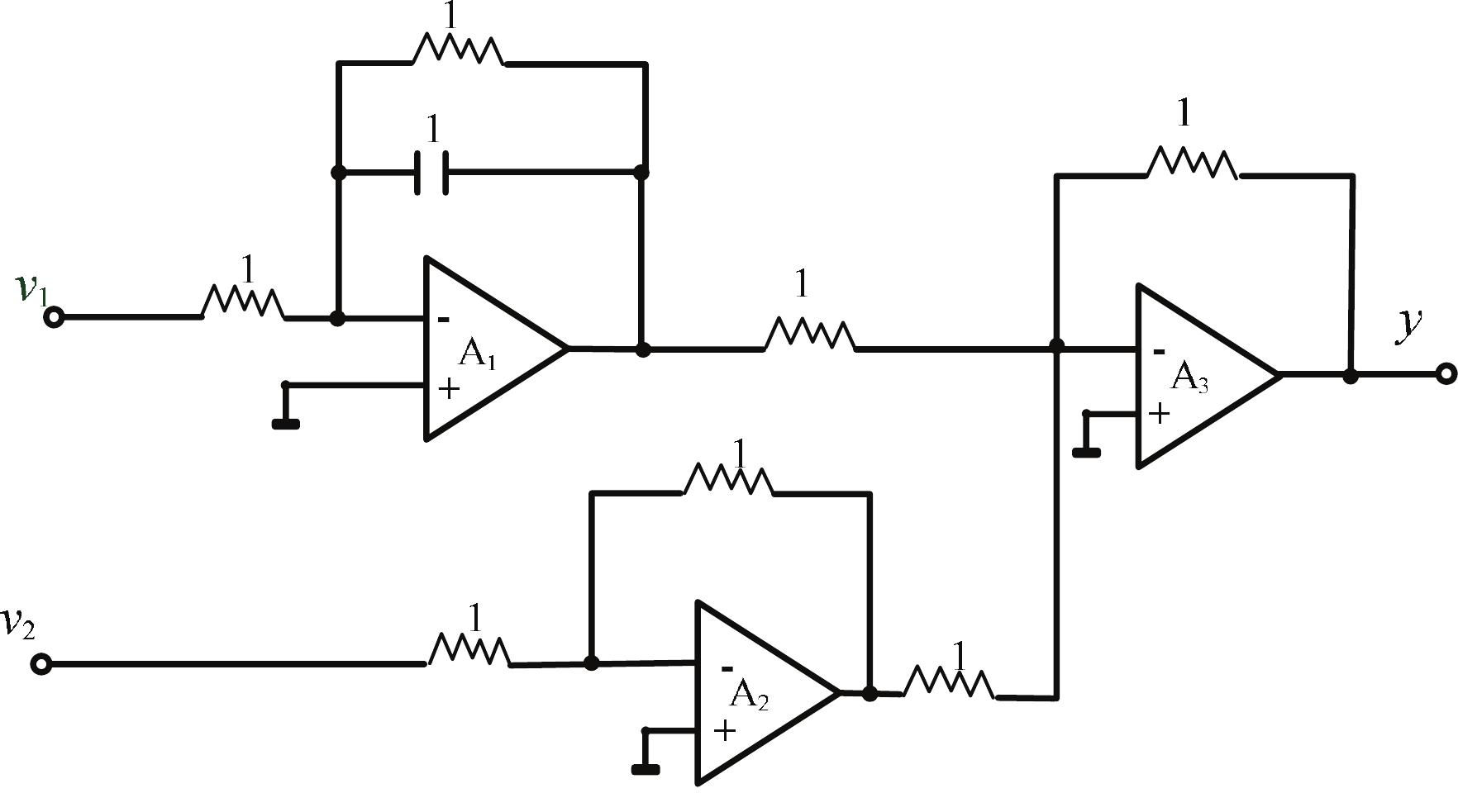}
\caption{Classical hardware implementation of the system
\eqref{into-example}. \label{fig:network-c}}
\end{figure}

With the birth and development of quantum technologies, quantum
control systems constructed using quantum optical devices play a
more and more important role  in  control engineering, 
\cite{WM93}, \cite{WM94b}, and \cite{WM09}. Linear systems are of
basic importance to control engineering, and also arise in the
modeling and control of quantum systems; see \cite{GZ00} and
\cite{WM09}.
A classical  linear system described by the state space
representation can be realized using electrical and electronic
components by linear electrical network synthesis theory, see
\cite{AV1973}. For example, consider a classical system given by
\begin{align}
d \xi(t)=&-\xi(t) dt +dv_{1}(t)\nonumber
\\
d y(t)=& \xi(t) dt + dv_{2}(t)\label{into-example}
\end{align}
where $\xi(t)$ is the state, $v_{1}(t)$ and $v_{2}(t)$ are inputs,
and $y(t)$ is the output. Implementation of  the system
\eqref{into-example} at the hardware level is  shown in Figure
\ref{fig:network-c}. Analogously to the electrical network synthesis
theory of how to synthesize linear analog circuits from basic
electrical components, \cite{NJD09} have proposed a quantum network
synthesis theory (briefly introduced in Subsection
\ref{subsect:quantum} of this paper), which details how to realize a
quantum system described by state space representations using
quantum optical devices.


\begin{figure}[htbp]
 \vspace{-10em}\centering
\includegraphics[scale=0.39]{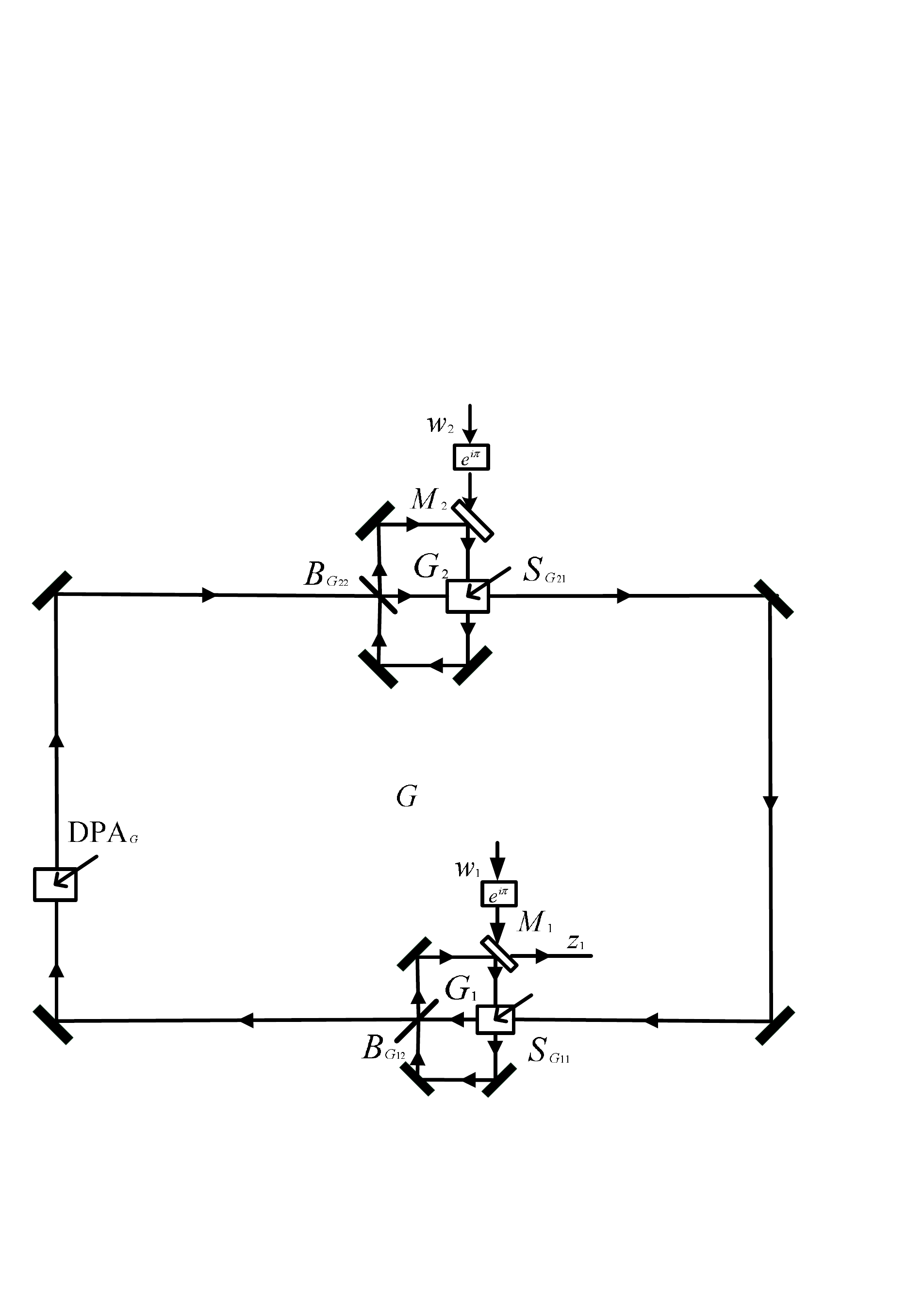}
\vspace{-6em}\caption{Quantum hardware realization of the system
\eqref{into-example}. \label{fig:realnew}}
\end{figure}

The purpose of this paper is  to address this issue of quantum
physical realization for a class of classical linear  systems. For
example, the quantum physical realization of the system
\eqref{into-example} is shown in Figure \ref{fig:realnew} (see
Example 1 in Section \ref{sec:main results} for more details). The
essential quantum optical components used in Figure
\ref{fig:realnew} include optical cavities, degenerate parametric
amplifiers (DPA), phase shifters, beam splitters, and squeezers,
etc;  interested readers may refer to \cite{BR04}, \cite{NJD09} for
a more detailed introduction to these optical devices. The problem
of quantum physical realization can be solved by embedding the
classical system into a larger linear quantum system, Theorem
\ref{thm:main}. In this way, the classical system is represented as
an invariant commutative subsystem of the larger quantum system.

While the results of this paper may be useful for a variety of
problems outside the scope of measurement feedback control, the
principle motivation for realizing classical systems in quantum
hardware is that one is better able to match the timescales and
hardware of a classical controller to the system being controlled.
Classical hardware is typically much slower than the quantum systems
intended to be controlled, and complex interface hardware may be
required. Compared with classical systems typically implemented
using standard analog or digital electronics, quantum mechanical
systems may provide a bandwidth much higher than that of
conventional electronics and thus increase processing times. For
instance, quantum optical systems can have frequencies up to
$10^{14}$ Hz or higher. Furthermore, it is becoming feasible  to
implement quantum networks in semiconductor materials, for example,
photonic crystals are periodic optical nanostructures that are
designed to affect the motion of photons in a similar way that
periodicity of a semiconductor crystal affects the motion of
electrons, and it may be desirable to implement control networks on
the same chip (rather than interfacing to a separate system); see
\cite{BKSWW}.

This paper is organized as follows. Section \ref{sec:preliminaries}
introduces some notations of classical and quantum random variables
and then gives a brief overview of classical linear  systems,
quantum linear stochastic systems as well as quantum  network
synthesis theory. Section \ref{sec:main results} presents the main
results of this paper, which are  illustrated with an example.
Section \ref{sec:application} presents a potential application of
the main results of Section \ref{sec:main results} to measurement
feedback control of quantum systems. Finally, Section
\ref{sec:conclusion} gives the conclusion of this paper.

{\em Notation.} The notations used in this paper are as follows:
$i=\sqrt{-1}$; the commutator is defined by $[A, B]=AB-BA$. If
$X=[x_{jk}]$ is a matrix of linear operators or complex numbers,
then $X^{\#} = [x^{*}_{jk}]$ denotes the operation of taking the
adjoint of each element of $X$, and $X^{\dag}=[x^{*}_{jk}]^T$ . We
also define $\Re(X)=(X + X^{\#})/2$ and $\Im(X)=(X -X^{\#})/2i$, and
$\mathrm{diag}_{n}(M)$ denotes a block diagonal matrix with a square
matrix $M$ appearing $n$ times on the diagonal block. The symbol
$I_n$ denotes the $n \times n$ identity matrix, and we write
\begin{align}
J_n = \left[ \begin{array}{cc} 0 & I_n
\\
-I_n & 0
\end{array} \right].
\end{align}
\section{Preliminaries}
\label{sec:preliminaries}

\subsection{Classical and quantum random variables}
\label{sec:preliminaries-random}

 Recall that a random variable $X$ is {\em
Gaussian} if its probability distribution $\mathbf{P}$ is Gaussian,
i.e.
\begin{equation}
\mathbf{P} ( a < X < b ) = \int_a^b p_X(x) dx,
\end{equation}
where $p_X(x) =
\frac{1}{\sigma\sqrt{2\pi}}\mathrm{exp}(-\frac{(x-\mu)^{2}}{2\sigma^2})$.
Here, $\mu = \mathbf{E}[ X]$ is the mean, and $\sigma^2 =
\mathbf{E}[ (X-\mu)^2 ]$ is the variance.

In quantum mechanics,  observables   are
 mathematical representations of physical quantities that
can (in principle) be measured, and state
 vectors $\psi$
summarize the status of physical systems and permit the calculation
of expected values  of observables.  State vectors may be described
mathematically as elements of a Hilbert space $\mathfrak{H}=
L^2(\mathbf{R})$ of square integrable complex-valued functions on
the real line, while observables are self-adjoint operators  $A$ on
$\mathfrak{H}$.  The expected value of an observable $A$ when in
pure state $\psi$ is given by the inner product  $\langle \psi, A
\psi \rangle=\int_{-\infty}^{\infty} \psi(q)^* A \psi(q)dq$.
Observables are {\em quantum random variables}.

A basic example is the quantum harmonic oscillator, a model for a
quantum particle in a potential well; see \cite[Chapter 14]{EM98}.
 The position and momentum of the particle
are represented by   observables $Q$ and $P$ (also called position quadrature and momentum quadrature),
respectively, defined by
\begin{equation}
(Q \psi)(q) = q \psi (q), \ \ \ (P\psi)(q) = -i    \frac{d}{dq}
\psi(q)
\end{equation}
for $\psi \in \mathfrak{H} = L^2(\mathbf{R})$. Here,   $q\in
\mathbf{R}$ represents  position values.  The position and momentum
operators do not commute, and in fact satisfy the commutation
relation $[Q,P]=i$.  In quantum mechanics, such non-commuting
observables are referred to as being incompatible. The state vector
\begin{equation}
\psi(q) =
(2\pi)^{-\frac{1}{4}}\sigma^{-\frac{1}{2}}\mathrm{exp}(-\frac{(q-\mu)^{2}}{4\sigma^2})
\end{equation}
is an instance of what is known as a {\em Gaussian state}. For this
particular {\em Gaussian state}, the means of $P$ and $Q$ are given by
$\int_{-\infty}^{\infty} \psi(q)^* Q\psi(q)dq =\mu$, and
$\int_{-\infty}^{\infty} \psi(q)^* P\psi(q)dq =0$, and similarly the
variances are $\sigma^2$ and $\frac{\hbar^2}{4\sigma^2}$,
respectively.

If we are given a classical  vector-valued random  variable
$\widetilde{X}=[X_1\quad X_2\quad\cdots\quad X_n]^T$, we may {\em
realize} (or represent) it using a quantum vector-valued random
variable $\check X_Q$ with associated state $\psi$  in a suitable
Hilbert space in the sense that the distribution of $\widetilde{X}$
is the same as the distribution of $\check X_Q$ with respect to the
state $\psi$. For instance, if the variable $\widetilde{X}$ have a
multivariate Gaussian distribution with its probability density
function given by
\begin{equation}
f(\tilde{x}) =
(2\pi)^{-\frac{n}{2}}|\Sigma|^{-\frac{1}{2}}\mathrm{exp}\left(-\frac{1}{2}(\tilde{x}-\tilde{\mu})^T\Sigma^{-1}(\tilde{x}-\tilde{\mu})\right)
\end{equation}
with mean $\tilde{\mu}\in\mathbb{R}^{n}$ and covariance matrix
$\Sigma\in\mathbb{R}^{n\times n}$, we may realize this classical
random variable $\widetilde{X}$ using an open harmonic oscillator.
Indeed, we can take the realization to be the position quadrature
$\check X_Q=[Q_1^T\quad Q_2^T\quad\cdots\quad Q_n^T]^T$ (for
example), with the state $\psi$ selected so that $(\tilde{\mu},
\Sigma^2)=( \tilde{\mu}_{Q}, \Sigma_{Q}^2)$. So statistically
$\widetilde{X} \equiv \check X_Q$.  The quantum vector $\check
X=[\check X_Q^T\quad \check X_P^T]^T$ is called an {\em
augmentation} of $\widetilde{X}$, where $\check X_P=[P_1^T\quad
P_2^T\quad\cdots\quad P_n^T]^T$ is the momentum quadrature. The
quantum realization of the classical random variable may be
expressed as
$
\widetilde{X }\equiv  \left[ \begin{array}{cc} I_n & 0_{n\times n}
\end{array} \right]
\left[  \begin{array}{c} \check X_Q\\ \check X_P
\end{array} \right].
$
\subsection{Classical linear  systems}
\label{sec:preliminaries-classical}

Consider a class of {\em classical} linear
systems of the form,
\begin{align}
d \xi (t)=&A\xi(t)dt +Bdv_1(t),\nonumber
\\
dy(t)=&C\xi(t) dt+ Ddv_2(t), \label{eq:classical1}
\end{align}
where $A\in\mathbb{R}^{n\times n}$, $B\in \mathbb{R}^{n\times
n_{v_1}}$, $C\in \mathbb{R}^{n_{y}\times n}$ and $D\in
\mathbb{R}^{n_{y}\times n_{v_2}}$ are real constant matrices,
$v_1(t)$ and $v_2(t)$ are input signals and independent.  The initial condition $\xi(0)=\xi_0$ is Gaussian,
while $y(0)=0$. The transfer function $\Xi_C(s)$ from the noise
input channel $v$ to the output channel $y$ for the classical system
(\ref{eq:classical1}) is denoted by
\begin{align}
\Xi_{C}(s)=&\left[\begin{tabular}{l|ll} $A$ & $\left[
         \begin{array}{cc}
           B, & 0_{n\times n_{v_2}} \\
         \end{array}
       \right]$\\
\hline \vspace{0em} $C$ & $\left[
         \begin{array}{cc}
          0_{n_{v_2}\times n_{v_2}}, & D \\
         \end{array}
       \right]$\\
\end{tabular}\right](s)=\left[
       \begin{array}{cc}
         C\left(sI_{n}-A\right)^{-1}B, & D \\
       \end{array}
     \right]
\end{align}

\subsection{Quantum  linear stochastic systems}
\label{sec:preliminaries-quantum}

Consider a {\em quantum} linear
stochastic system of the form  (see  e.g.
\cite{GZ00},   \cite{WM09}, \cite{NJP09})
\begin{align} \label{eq:quantum1}
dx(t)=&\widetilde{A}x(t)dt+\widetilde{B}dw(t),\nonumber
\\
dz(t)=&\widetilde{C}x(t)dt+\widetilde{D}dw(t),
\end{align}where   $\widetilde{A}\in\mathbb{R}^{2n\times 2n}$, $\widetilde{B}\in\mathbb{R}^{2n\times n_w}$, $\widetilde{C}\in\mathbb{R}^{n_z\times 2n}$ and $\widetilde{D}\in\mathbb{R}^{n_z\times n_w}$ are real constant matrices. We assume that $n_w$ and $n_z$ are even, with $n_z \leq n_w$ (see \cite[Section II]{JNP08} for details). We refer to $n$ as the {\em degrees of freedom} of systems of the form (\ref{eq:quantum1}).
 Equation (\ref{eq:quantum1}) is a {\em quantum stochastic differential equation} (QSDE) \cite{KRP92} and \cite{GZ00}.
  In equation (\ref{eq:quantum1}), $x(t)$ is a vector of self-adjoint possibly non-commuting operators, with the initial value $x(0)=x_0$
satisfying the commutation relations
\begin{equation}
x_{0j} x_{0k} - x_{0k} x_{0j} = 2 i \widetilde \Theta_{jk},
\label{eq:CCR}
\end{equation}
where $\widetilde \Theta=[\widetilde \Theta_{jk}]_{j,k=1,2,\ldots,2n}$ is a skew-symmetric real matrix.  The
matrix $\widetilde \Theta$ is said to be {\em canonical} if it is
the form $\widetilde \Theta = J_{n}$. The components of the vector
$w(t)$ are quantum stochastic processes with the following non-zero
Ito products:
\begin{equation}
dw_j(t) dw_k(t) = F_{jk}dt,
\end{equation}
where $F$ is a non-negative definite Hermitian matrix.  The matrix $F $ is
said to be {\em canonical} if it is the form $F = I_{n_w} +
iJ_{\frac{n_w}{2}}$. In this paper we will take $\widetilde \Theta$ and $F$ to
be canonical. The transfer function for the quantum linear stochastic system
(\ref{eq:quantum1}) is given by
 \vspace{-2em} {\small\begin{align}
\Xi_{Q}(s)=\left[\begin{tabular}{l|ll}
$\widetilde{A}$ & $\widetilde{B}$\\
\hline \vspace{-1em}\\
$\widetilde{C}$ & $\widetilde{D}$\\
\end{tabular}\right](s)=\widetilde{C}\left(sI_{2n}-\widetilde{A}\right)^{-1}\widetilde{B}+\widetilde{D}.
\end{align}}Here we
mention that while the equations (\ref{eq:quantum1}) look formally
like the classical equations (\ref{eq:classical1}), they are not
classical equations, and in fact give the Heisenberg dynamics of a
system of coupled open quantum harmonic oscillators. The variables
$x(t)$, $w(t)$ and $z(t)$ are in fact vectors of quantum observables
(self-adjoint non-commuting operators, or quantum stochastic
processes).

The quantum system (\ref{eq:quantum1}) is {\em (canonically)
physically realizable} (cf.\cite{WNZJ2012}), if and
only if the matrices $\widetilde{A}$, $\widetilde{B}$,
$\widetilde{C}$ and  $\widetilde{D}$ satisfy the following
conditions:
\begin{eqnarray}
&&\widetilde{A} J_n + J_n \widetilde{A}^{T}+\widetilde{B} J_{\frac{n_w}{2}}  \widetilde{B}^{T}=0 ,
\label{condition1}\\
&&\widetilde{B}J_{\frac{n_w}{2}}\widetilde{D}^T= -J_n \widetilde{C}^T,\label{condition2}\\
&&\widetilde{D}J_{\frac{n_w}{2}}\widetilde{D}^T=J_{\frac{n_z}{2}}. \label{condition3}
\end{eqnarray}
where $n_w\geq n_z$. In fact, under these conditions   the quantum linear stochastic
system (\ref{eq:quantum1})  corresponds to an open quantum harmonic
oscillator \cite[Theorem 3.4]{JNP08}  consisting of $n$
oscillators (satisfying  canonical commutation relations) coupled to
$n_w/2$ fields (with canonical Ito products and commutation
relations). In particular, in the canonical case,
$x_0=(q_1,q_2,\ldots,q_n,p_1,p_2,\ldots,p_n)^T$, where $q_j$ and
$p_j$ are the position and momentum operators of the oscillator $j$
(which constitutes the $j$th of degree of freedom of the system)
that satisfy the commutation relations $[q_j,p_k]=2i\delta_{jk}$,
$[q_j,q_k]=[p_j,p_k]=0$ in accordance with (\ref{eq:CCR}). Hence by
the results of \cite{NJD09} the system can be implemented using
standard quantum optics components. It is also possible to consider
other quantum physical implementations.

\subsection{Quantum  network synthesis theory}\label{subsect:quantum}
We briefly review some definitions and results from \cite{NJD09};
see also \cite{HIN2010a} and \cite{HIN2010b}. The quantum linear
stochastic system (\ref{eq:quantum1}) can be reparametrized in terms
of three parameters $S,L,H$ called the scattering, coupling and
Hamiltonian operators, respectively.  Here $S$ is a complex unitary $\frac{n_w}{2} \times \frac{n_w}{2}$
matrix $S^{\dag}S= S S^{\dag}=I$,  $L=\Lambda x_0$ with $\Lambda \in
\mathbb{C}^{ \frac{n_w}{2} \times 2n}$, and $H=\frac{1}{2}x_0^T R x_0$ with
$R=R^T \in \mathbb{R}^{2n \times 2n}$.  Recall that there is a
one-to-one correspondence between the matrices $\tilde A, \tilde B,
\tilde C, \tilde D$ in (\ref{eq:quantum1}) and the triplet $S,L,H$
or equivalently the triplet $S,\Lambda,R$; see \cite{JNP08} and
\cite{NJD09}. Thus, we can represent a quantum linear stochastic
system $G$ given by (\ref{eq:quantum1}) with the shorthand notation
$G=(S, L,H)$ or $G=(S,\Lambda,R)$ \cite{GJ09a}. Given two quantum
linear stochastic systems $G_{1}=(S_{1},L_{1},H_{1})$ and
$G_{2}=(S_{2},L_{2},H_{2})$ with the same number of field channels,
the operation of cascading of $G_{1}$ and $G_{2}$ is represented by
the series product $G_{2}\triangleleft G_1$ defined by
 \vspace{-2em}{\small\begin{equation*} G_{2}\triangleleft G_{1}=\biggl(S_{2}S_{1},
L_{2}+S_{2} L_{1}, H_{1}+H_{2}\\+\frac{1}{2i}(L_{2}^{\dagger}S_{2}
L_{1}-L_{1}^{\dagger}S_{2}^{\dagger}L_{2})\biggl)
\end{equation*}}
According to \cite[Theorem 5.1]{NJD09} a linear quantum stochastic
system with $n$ degrees of freedom can be decomposed into an
unidirectional connection of $n$ one degree of freedom harmonic
oscillators with a direct coupling between two adjacent one degree
of freedom quantum harmonic oscillators. Thus an arbitrary quantum
linear stochastic system  can in principle be synthesized if:

1) Arbitrary one degree of freedom systems of the form
(\ref{eq:quantum1}) with $n_{w}$ input fields and $n_{w}$ output
fields can be synthesized.

2) The bidirectional coupling
$H^{d}=\sum^{n-1}_{j=1}\sum^{n}_{k=j+1}x_{k}^{T}\times\\\left(R_{jk}^{T}-
\frac{1}{2i}(\Lambda_{k}^{\dagger}\Lambda_{j}-\Lambda_{k}^{T}\Lambda_{j}^{\#})\right)x_{j}$
can be synthesized, where $\Lambda_j$ denotes the $j$th row of the
complex coupling matrix $\Lambda$ . The Hamiltonian matrix $R$ is
given by
$R=\frac{1}{4}P_{2n}^{T}(-J_{n}\widetilde{A}+\widetilde{A}^{T}J_{n})P_{2n}$
and the coupling matrix $\Lambda$ is given by
$\Lambda=-\frac{i}{2}\left[
                      \begin{array}{cc}
                        0_{n_{w}\times
n_{w}}& I_{n_{w}} \\
                      \end{array}
                    \right]
P_{2n_{w}}\mathrm{diag}_{n_{w}}(M)P_{2n_{w}}^{T}\widetilde{B}^{T}J_{n}P_{2n}$
where $M=\left[
           \begin{array}{cc}
             1 & -i \\
             1 & i \\
           \end{array}
         \right]
$, $P_{2n}$ denotes a permutation matrix acting on a column vector
$f=[f_1\quad f_2\quad ...\quad f_{2n}]^{T}$ as $P_{2n}f$=$[f_1\quad
f_{1+n}\quad f_2\quad f_{2+n}\quad ...\quad f_n\quad f_{2n}]^T$.

The work \cite{NJD09} then shows how  one degree of freedom
systems and the coupling $H^d$ can be approximately implemented
using certain linear and nonlinear quantum optical components. Thus
in principle any system of the form (\ref{eq:quantum1}) can be
constructed using these components.  In Section \ref{sec:main results}
we will use the construction proposed in \cite{NJD09} to realize
systems of the form (\ref{eq:quantum1}) without further comment. The
details of the construction and the individual components involved
can be found in \cite{NJD09} and the references therein.

\section{Quantum Physical Realization}\label{sec:main results}
\begin{figure}[h]
\begin{center}
\includegraphics[scale=0.7]{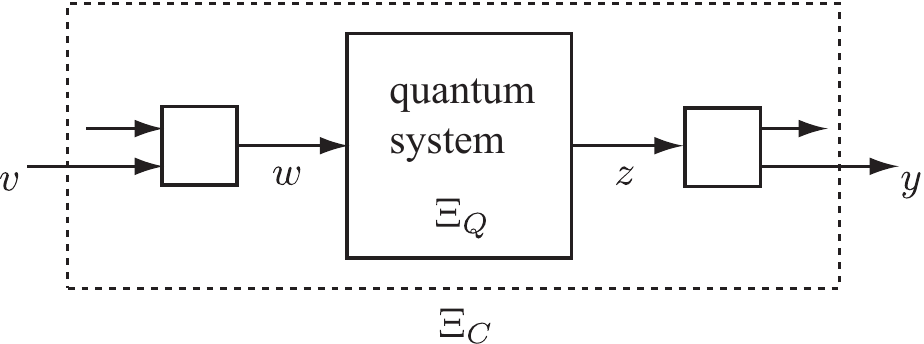} 
\end{center}
\caption{Quantum realization of classical system $\Xi_C : v \mapsto
y$.} \label{fig:q-realize1}
\end{figure}

In this section we present our results
concerning the quantum physical realization of classical linear
systems and then provide an example to illustrate the results.
As is well known, for a linear system, its state space representation can be
associated to a unique transfer function representation. Then, we will  show how the transfer function matrix
$\Xi_C(s)$ can be realized (in a sense to be defined more precisely
below) using linear {\em quantum} components. In general, the dimension of vectors in (\ref{eq:quantum1}) is greater than the vector dimension in (\ref{eq:classical1}), and so to obtain a quantum realization of the classical system  (\ref{eq:classical1}) using the quantum system  (\ref{eq:quantum1}) we require that the transfer functions be related by
 \begin{equation}
\Xi_C(s) = M_o \Xi_Q(s) M_i, \label{eq:realize-0}
\end{equation}
as illustrated in Figure \ref{fig:q-realize1}. Here, the matrix
$M_i$ and $M_o$ correspond to operation of selecting elements of the
input vector $w(t)$ and the output vector $z(t)$ of the quantum
realization that correspond to  quantum representation of $v(t)$ and
$y(t)$, respectively (as discussed in Section
\ref{sec:preliminaries}). In Figure \ref{fig:q-realize1}, the
unlabeled box on the left indicates that $v(t)$ is represented as
some subvector  of  $w(t)$ (e.g. modulation\footnote[1]{Modulation
is the process of merging two signals to form a third signal with
desirable characteristics of both in a manner suitable for
transmission.}), whereas the unlabeled box on the right indicates
that $y(t)$ corresponds to  some subvector  of $z(t)$ (quadrature
measurement).

\begin{definition}    \label{dfn:realize1}
The classical linear stochastic system (\ref{eq:classical1}) is said
to be canonically {\em realized} by the quantum linear stochastic
system (\ref{eq:quantum1}) provided:
\begin{enumerate}

\item The dimension of the quantum vectors $x(t)$, $w(t)$ and $z(t)$ are
twice the lengths of the corresponding classical vectors $x(t)$,
$v(t) = [v_1(t)^T \quad  v_2(t)^T ]^T$ and $y(t)$, where
$x(t)$=$[\xi(t)^T\quad\theta(t)^T]^T$ with $\xi(t)=[q_1(t)\quad
q_2(t)\quad \cdots \quad q_n(t)]^T$ and $\theta(t)=[p_1(t)\quad
p_2(t)\\\quad\cdots \quad p_n(t)]^T$, $w(t)=[v_1(t)^T\quad
v_2(t)^T\quad u_1(t)^T\quad u_2(t)^T]^T$ and $z(t)=[y(t)^T\quad
y'(t)^T]^T$.

\item The classical
$\Xi_C(s)$ and quantum $\Xi_Q(s)$ transfer functions are related by
equation (\ref{eq:realize-0}) for the choice
\begin{equation*}
M_o=\left[\begin{array}{cc}
                              I_{n_{y}} & 0_{n_{y}\times n_{y}}  \\
                            \end{array}
                          \right],\quad
M_i=\left[\begin{array}{cc}
                             I_{n_v} &  0_{n_v\times n_v} \\
                            \end{array}\right]^T.
\end{equation*}
\item The quantum linear stochastic system (\ref{eq:quantum1}) is
canonically physically realizable (as described in Section
\ref{sec:preliminaries-quantum})  with the system matrices $\tilde
A, \tilde B, \tilde C$ and $\widetilde{D}$ having the special structure:
\begin{eqnarray}\label{matrices}
\!\!\!\!\!\!\!\!\!\!&&\!\!\!\!\!\!\!\!\!\!\widetilde{A} = \left[ \begin{array}{cc} A_0 & 0_{n\times n}
\\
A_1 & A_2
\end{array} \right]\!\!,
\widetilde{B} = \left[ \begin{array}{cccc} B_0 & 0_{n \times
n_{v_2}} & 0_{n \times n_{v_1}} & 0_{n \times n_{v_2}}
\\
B_1 & B_2 & B_3 & B_4
\end{array} \right],\nonumber\\
\!\!\!\!\!\!\!\!\!\!&&\!\!\!\!\!\!\!\!\!\!\widetilde{C} = \left[ \begin{array}{cc} C_0 & 0_{n_w \times n}
\\
C_1 & C_2
\end{array} \!\!\right]\!\!, \widetilde{D} = \left[ \begin{array}{cccc}  0_{n_y \times n_{v_1}} & D_0& 0_{n_y \times n_{v_1}} & 0_{n_y \times n_{v_2}}
\\
D_1 & D_2 & D_3 & D_4
\end{array} \!\!\right]\!\!.
\end{eqnarray}
with $A_0 \in \mathbb{R}^{n \times n}$, $A_{1}\in
\mathbb{R}^{n\times n}$, $A_{2}\in \mathbb{R}^{n\times n}$, $B_0 \in
\mathbb{R}^{n \times n_{v_1}}$, $B_1 \in \mathbb{R}^{n \times
n_{v_1}}$, $B_{2}\in \mathbb{R}^{n\times n_{v_2}}$, $B_{3}\in
\mathbb{R}^{n\times n_{v_1}}$, $B_{4}\in \mathbb{R}^{n\times n_{
v_2}}$, $C_0\in \mathbb{R}^{n_y\times n}$, $C_{1}\in
\mathbb{R}^{n_y\times n}$, $C_{2}\in \mathbb{R}^{n_y\times n}$, $D_0
\in \mathbb{R}^{n_y \times n_{v_2}}$, $D_1 \in \mathbb{R}^{n_y
\times n_{v_1}}$, $D_{2}\in \mathbb{R}^{n_y\times n_{v_2}}$,
$D_{3}\in \mathbb{R}^{n_y\times n_{v_1}}$, and $D_{4}\in
\mathbb{R}^{n_y\times n_{ v_2}}$.
\end{enumerate}
\end{definition}
\begin{remark}
According to the structure of the matrices $\widetilde{A}$,
$\widetilde{B}$, $\widetilde{C}$, and $\widetilde{D}$, and since the system  (\ref{eq:quantum1}) is physically realizable, it can be
verified directly that commutation relations for $\xi(t), \theta(t)$
satisfy $[\xi(t), \xi(s)^T]=0$, $[\xi(t), \theta(s)^T]\neq0$ and
$[\theta(t), \theta(s)^T]=0$ for all $t, s \geq 0$. The quantum
realization of the classical variable $\xi(t)$ may be expressed as
$\xi(t)=\left[ \begin{array}{cc} I & 0
\end{array} \right]x(t)=\left[ \begin{array}{cc} I & 0
\end{array} \right]
\left[  \begin{array}{c}\xi(t)  \\ \theta(t)
\end{array} \right]$. The structures of the matrices $\tilde A, \tilde B,\tilde C$ and
$\widetilde{D}$ in the above definition ensure that the classical
system (\ref{eq:classical1}) can be embedded as an invariant
commutative subsystem of the quantum system (\ref{eq:quantum1}), as
discussed in \cite{JNP08}, \cite{GJ09a} and \cite{WNZJ2012}. Here,
the classical  variables and the classical signals are represented
within an invariant commutative subspace of the full quantum
feedback system, and the additional quantum degrees of freedom
introduced in the quantum controller have no influence on the
behavior of the feedback system; see \cite{JNP08} for details. In
fact, $\widetilde{D}$ represents static Bogoliubov transformations
or symplectic transformations, which can be realized as a suitable
static quantum optical network (eg. ideal squeezers), \cite{NJD09},
  \cite{Nurdin2012}.
\end{remark}

In what follows we restrict our attention to stable classical
systems, since it may not be desirable to attempt to implement an
unstable quantum system. By a {\em stable} quantum system
(\ref{eq:quantum1}) we mean that the $\tilde A$ is Hurwitz.  We will seek  stable
quantum realizations. Furthermore, given the quantum physical
realizability conditions \eqref{condition1}-\eqref{condition3}, we cannot do the quantum
realizations for an arbitrary classical system (\ref{eq:classical1}). For these reasons we make the following
assumptions regarding the classical linear stochastic system
(\ref{eq:classical1}).
\begin{assumption}\label{assume:classical1}
Assume the following conditions hold:
\begin{enumerate}
\item The matrix $A$ is a Hurwitz matrix.
\item The pair $(-A, B)$ is stabilizable.
\item The matrix $D$ is of full row rank.
\end{enumerate}
\end{assumption}
\begin{theorem} \label{thm:main}
Under Assumption \ref{assume:classical1}, there exists a stable quantum linear
stochastic system (\ref{eq:quantum1}) realizing  the given classical
linear stochastic system (\ref{eq:classical1}) in the sense of
Definition \ref{dfn:realize1}, where the matrices $\tilde A,\tilde
B, \tilde C$ and $\tilde D$ can be constructed according to the
following steps:
\begin{enumerate}

\item $A_0=A$, $B_0=B$, $C_0=C$ and $D_0=D$, with $A$ $B$, $C$ and $D$ as given in (\ref{eq:classical1}).

\item
$B_1$, $B_2$  are arbitrary matrices of suitable
dimensions.

\item
The matrices $A_2$ and $B_3$ can be fixed simultaneously by
\begin{equation}
 A_2=-A^T-B_3B^T
\label{A2}
\end{equation}
where $B_3$ is chosen to let $A_2$ be a Hurwitz matrix.

\item
The matrices $B_4$ and $D_4$ are given by
\begin{align}
&B_4=-C^{T}(DD^T)^{-1}D+N_1(D)^T\label{B2},\\
&D_4=(DD^T)^{-1}D+N_2(D)^T,\label{D3}
\end{align}
where $N_1(D)$ (resp., $N_2(D))$ denotes a matrix of the same dimension as $B_4^T$ (resp., $D_4^T$) whose columns are in the kernel space of $D$.

\item For a given $D_4$, there always  exist matrices $D_1,D_2,D_3$ satisfying
\begin{equation}
 -D_3D_1^T - D_4D_2^T + D_1D_3^T + D_2D_4^T =0.
\end{equation}
The simplest choice is $D_1=0$, $D_2=0$, and $D_3=0$.

\item
The remaining matrices can be constructed as follows,
\begin{align}\label{D1D4}
C_2=-D_3B^T&& \\
C_1=D_4B_2^T+D_3B_1^T-D_2B_4^T-D_1B_3^T&& \\
A_1=\Xi+\frac{1}{2}(B_3B_1^T-B_1B_3^T-B_2B_4^T+B_4B_2^T)&&
\label{A1}
\end{align}
where $\Xi$ is an arbitrary $n\times n$ real symmetric matrix.
\end{enumerate}
\end{theorem}
\noindent \textbf{Proof.} The idea of the proof is to represent the
classical stochastic processes $\xi(t)$ and
$v(t)$ as quadratures of quantum stochastic
processes $x(t)$ and $w(t)$ respectively, and then determine the
matrices $\widetilde{A}$, $\widetilde{B}$, $\widetilde{C}$ and
$\widetilde{D}$ in such a way that the requirements of Definition
\ref{dfn:realize1} and the Hurwitz property of $\widetilde A$ are
fulfilled. To this end, we set the number of oscillators to be
$n=n_c$, the number of field channels as $n_w =
2n_v=2(n_{v_1}+n_{v_2})$ and the number of output field channels as $n_z=2n_{y}$.
Equations (\ref{A2})-(\ref{A1}) can be obtained from the physical
realizability constraints (\ref{condition1})-(\ref{condition3}).
According to the second assumption of Assumption
\ref{assume:classical1}, we can choose $B_3$ such that
$A_2=-A^T-B_3B^T$ is a Hurwitz matrix. From the first assumption of
Assumption \ref{assume:classical1}, we can conclude that
$\widetilde{A}$ is a Hurwitz matrix, which means the quantum linear
stochastic system (\ref{eq:quantum1}) is stable. Using $M_i$ and
$M_o$ as defined in Definition \ref{dfn:realize1} and then combining
these with equations (\ref{matrices})-(\ref{A1}), we can verify the
following relation between the classical $\Xi_C(s)$ and quantum
$\Xi_Q(s)$ transfer functions,{\setlength\arraycolsep{3pt}
\begin{align*} &M_o \Xi_Q(s) M_i
\\ =&\left[\begin{array}{cc}
                              I_{n_{y}} & 0_{n_{y}\times n_{y}}  \\
                            \end{array}
                          \right]\biggl\{\left[ \begin{array}{cc} C &0_{n_y \times n}
\\
C_1 & C_2
\end{array} \right]\left(sI_{2n}-\left[ \begin{array}{cc} A & 0_{n\times n}
\\
A_1 & A_2
\end{array} \right]\right)^{-1}\!\!\!\!\times \nonumber\\&\left[ \begin{array}{cccc} B & 0_{n \times n_{v_2}} & 0_{n \times
n_{v_1}} & 0_{n \times n_{v_2}}
\\
B_1 & B_2 & B_3 & B_4
\end{array} \right]+\widetilde{D}\biggr\}\left[
                            \begin{array}{cc}
                             I_{n_v} &  0_{n_v\times n_v} \\
                            \end{array}
                          \right]^T
\\ =&\left[ \begin{array}{cc} C & 0_{n_{y}\times n}
\end{array} \right]\left[ \begin{array}{cc} (sI_{n}-A)^{-1}& 0_{n\times n}
\\
(sI_{n}-A_2)^{-1}A_1(sI_{n}-A)^{-1} &(sI_{n}-A_2)^{-1}
\end{array} \right]\\&\times\left[ \begin{array}{cc} B& 0_{n \times n_{v_2}}
\\
B_1&B_2
\end{array} \right]+\left[\begin{array}{cc}
                                       0_{n_{y}\times n_{v_{1}}}& D \\
                                    \end{array}
                                  \right]
\\=&\left[C\left(sI_{n}-A\right)^{-1}B \quad \quad D
     \right]=\Xi_C(s)
\end{align*}}
This completes the proof.\hfill $\square \ $\vspace{-3mm}

$\mathbf{Example \ 1}$: \ Let us realize the classical system
\eqref{into-example} introduced in Section \ref{sec:intro}. The
classical transfer function is
$\Xi_{C}(s)=\left[\begin{array}{cc}\frac{1}{s+1} & 1
\\\end{array}\right]$. By Theorem \ref{thm:main}, we can construct a quantum system $G$
given by
\begin{eqnarray}
dx_1 &=& -x_1dt + dv_1
\nonumber \\
dx_2 &=&   -x_2 dt  + 2 du_1   -  du_2
\nonumber \\
dz_1 &=& x_1 dt + dv_2
\nonumber \\
dz_2 &=&  du_2
\label{eq:eg1}
\end{eqnarray}
The quantum transfer function is given by
$
\Xi_{Q}(s) = \left[ \begin{array}{cccc}
\frac{1}{s+1} & 1 & 0 & 0
\\
0  & 0 & 0  & 1
\end{array}
\right].
$ Since in this case $M_o=\left[ \begin{array}{cc}
                              1 & 0   \\
                            \end{array}
                          \right]
$, $M_i=\left[\begin{array}{cc}
                             I_{2} &  0_{2\times 2} \\
                            \end{array}\right]^T$, we  see that $\Xi_C(s) =M_o \Xi_Q(s)
                            M_i$.
The commutative subsystem $d x_1 = - x_1 dt + dv_1$, $dz_1 = x_1 dt  + dv_2$ can clearly be seen in these equations,
with the identifications
$y= z_1$,  $\xi=x_1$. It can be seen that $\widetilde{A}$,
$\widetilde{B}$, $\widetilde{C}$ and $\widetilde{D}$ satisfy the
physically realizable constraints (\ref{condition1}) and
(\ref{condition2}).

Let us realize this classical system. The parameter $R$ for $G$ is given by $R=0$, 
    which means no Degenerate Parametric Amplifier (DPA) is required to implement $R$; see \cite[section 6.1.2]{NJD09}. The coupling matrix $\Lambda$ for $G$ is given by
{\setlength\arraycolsep{1.5pt}
\begin{equation*}
\Lambda =\left[
                \begin{array}{c}
                  \Lambda_{1} \\
                 \Lambda_{2} \\
                \end{array}
              \right]=\left[\begin{array}{cc}
                                      -1&  -0.5i \\
                                      0.5& 0\\
                                    \end{array}
                                  \right]
\end{equation*}}
From the above equation, we can get $\Lambda_{1}=[\begin{array}{cc} -1 & -0.5i \end{array}]$ and $\Lambda_{2}=[\begin{array}{cc} 0.5 & 0 \end{array}]$. The coupling operator
$L_{1}=\Lambda_1 x_0$ for $G$ is given by
\begin{align}\label{l11}
L_{1}=\Lambda_{1}\left[\begin{array}{c}q\\
                                    p \\
                                  \end{array}
                              \right]=\Lambda_{1}\left[
                  \begin{array}{cc}
                    1 & 1 \\
                    -i & i \\
                  \end{array}
                \right]\left[\begin{array}{c}a\\
                                    a^{*} \\
                                  \end{array}
                              \right]=-1.5a-0.5a^*
                              \end{align}
where $a=\frac{1}{2}(q+ip)$ is the oscillator annihilation operator
and $a^{*}=\frac{1}{2}(q-ip)$ is the creation operator of  the
system $G$ with position and momentum operators $q$ and $p$,
respectively. $L_{1}$ can be approximately realized by the
combination of a two-mode squeezer $\Upsilon_{G_{11}}$, a beam
splitter $B_{G_{12}}$, and an auxiliary cavity  $G_{1}$. If the
dynamics of $G_1$ evolve on a much faster time scale than that of
$G$ then the coupling operator $L_{1}$ is approximately given by:
$L_{1}=\frac{1}{\sqrt{\gamma_{1}}}(-\epsilon_{12}^{*}a+\epsilon_{11}a^{*})$,
where $\gamma_1$ is the coupling coefficient of the only partially
transmitting mirror of $G_1$, $\epsilon_{11}$ is the effective pump
intensity of $\Upsilon_{G_{11}}$ and $\epsilon_{12}$ is the
coefficient of the effective Hamiltonian for $B_{G_{12}}$ given by
$\epsilon_{12}=2\Theta_{12}e^{-i\Phi_{12}}$, where $\Theta_{12}$ is
the mixing angle of $B_{G_{12}}$ and $\Phi_{12}$ is the relative
phase between the input fields introduced by $B_{G_{12}}$; see
\cite{NJD09}. For this to be 
 a good approximation we require that
$\sqrt{\gamma_1}, |\epsilon_{11}|,|\epsilon_{12}|$  be sufficiently
large, and assuming that the coupling coefficient of the
mirror $M_{1}$ is $\gamma_{1}=100$, then we can get
$\epsilon_{11}=-5$, $\epsilon_{12}=15$, $\Phi_{12}=0$ and
$\Theta_{12}=7.5$. The scattering matrix for $G_1$ is $e^{i\pi}=-1$
and all other parameters are set to $0$.
In a similar way, the coupling operator $L_{2}=\Lambda_2 x_0$ can be
realized by the combination of $\Upsilon_{G_{21}}$, $B_{G_{22}}$,
and $G_{2}$. In this case, if we set the coupling coefficient of the
partially transmitting mirror  $M_{2}$ of $G_2$ to $\gamma_{2}=100$,
we find the effective pump intensity $\epsilon_{21}$ of
$\Upsilon_{G_{21}}$ given by $\epsilon_{21}=5$, the relative phase
$\Phi_{22}$ of  $B_{G_{22}}$ given by $\Phi_{22}=\pi$, the mixing
angle $\Theta_{22}$ of $B_{G_{22}}$ given by $\Theta_{22}=2.5$, the
scattering matrix for $G_2$ to be  $e^{i\pi}=-1$, with all other
parameters set to $0$. The implementation of the quantum system $G$
is shown in Figure \ref{fig:realnew}.

\section{Application}
\label{sec:application}

\begin{figure}[!htp]
\vspace{-4em}
    \centering
      \includegraphics[scale=0.4]{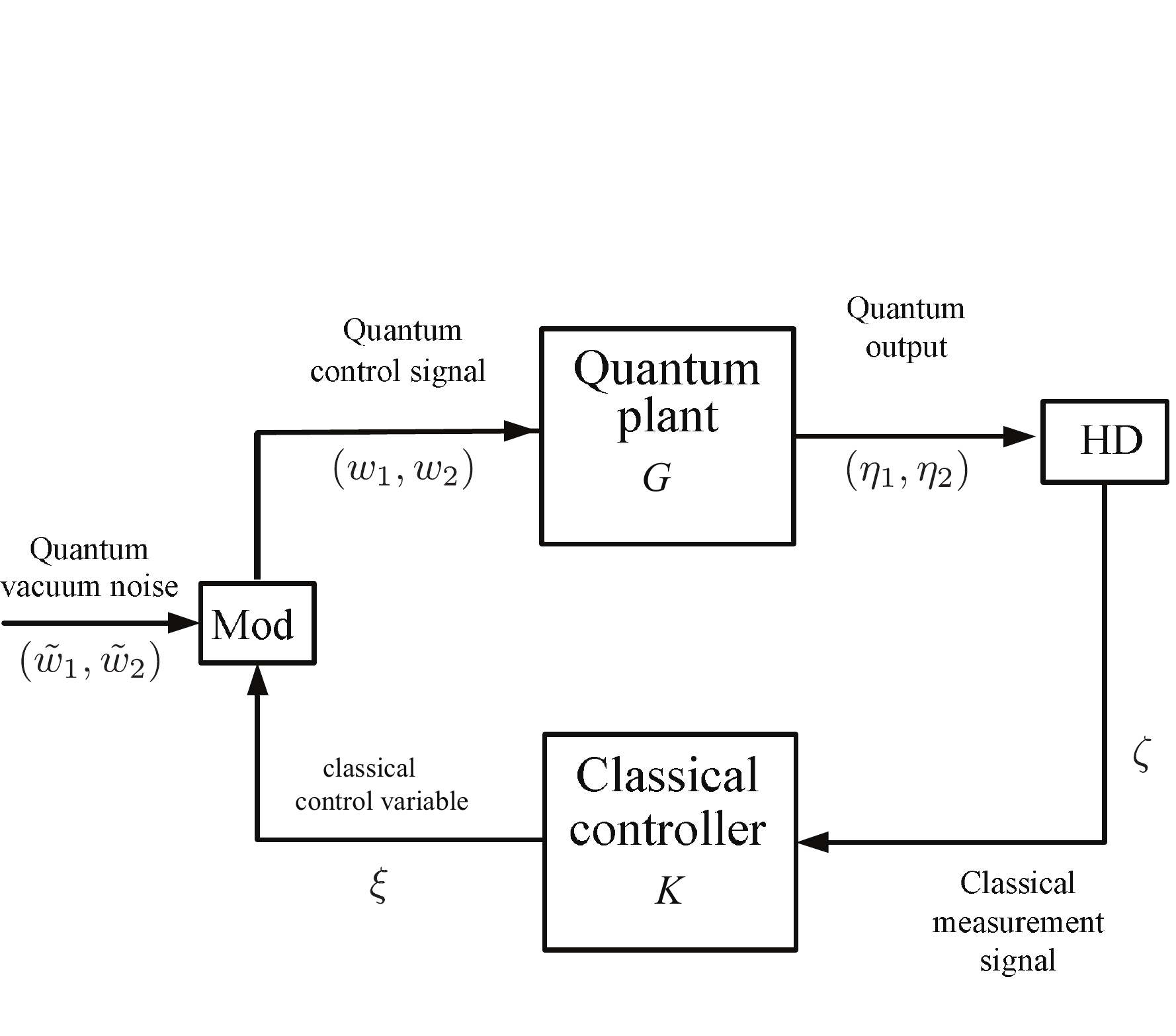}
\vspace{-2em}
\caption{Measurement feedback control of a quantum system, where HD represents the homodyne detector and Mod represents the optical modulator.
}\label{fig:q-c-loop1}
\end{figure}

The main results of this paper may have a practical application in measurement feedback control of quantum systems, which is important in  a
number of areas of quantum technology,  including quantum optical
systems, nanomechanical systems, and circuit QED systems; see
  \cite{WM93} and \cite{WM09}. In measurement feedback
control, the plant is a quantum system, while the controller is a
classical  system \cite{WM09}. The classical controller
processes the outcomes of a measurement of an observable of the
quantum system   to
determine the classical control actions that are applied to control
the behavior of the quantum system. The closed loop system involves
both quantum and classical components, such as an electronic device
for measuring a quantum signal, as shown in Figure
\ref{fig:q-c-loop1}. However, an important practical problem for the implementation of
measurement feedback control systems in Figure
\ref{fig:q-c-loop1} is the relatively slow speed of standard
classical electronics.

\begin{figure}[h]
    \begin{center}
     \includegraphics[scale=0.25]{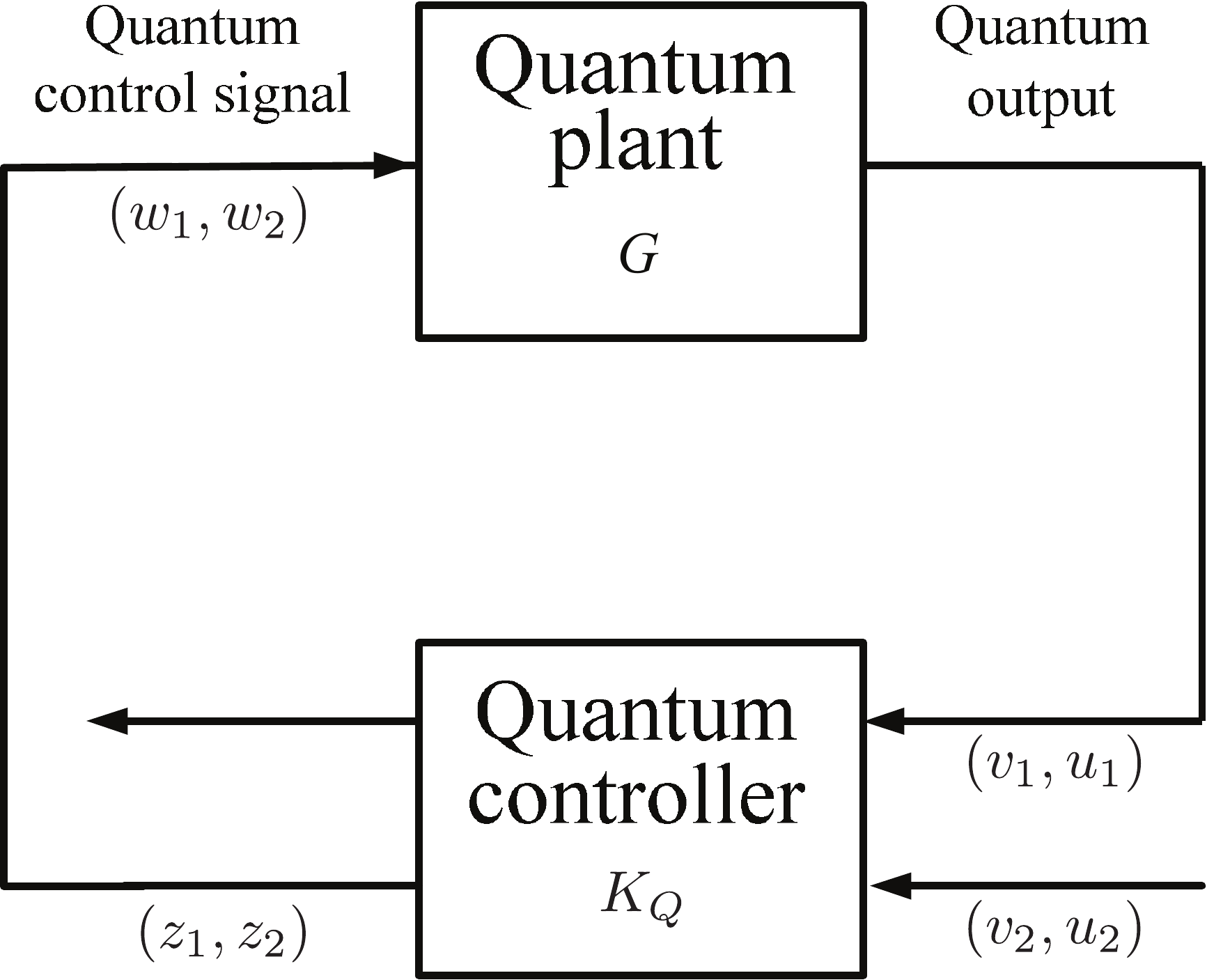} 
\end{center}
\caption{Quantum realization of a measurement feedback control
system.} \label{fig:q-c-loop2}
\end{figure}

According to the main results of Section \ref{sec:main results},  it may be possible to
realize the measurement feedback loop illustrated in Figure \ref{fig:q-c-loop1} fully at the quantum level. For instance, if the plant is a quantum optical
system where the classical control is a signal modulating a laser
beam, and if the measurement of the plant output (a quantum field)
is a quadrature measurement (implemented by a homodyne detection
scheme), then the closed loop system might be implemented fully
using quantum optics, Figure \ref{fig:q-c-loop2}.
The quantum implementation of the controller is designed so that (i)  its dynamics depend only on the required quadrature of the field (the quadrature that was measured in Figure \ref{fig:q-c-loop1}), and (ii) its output field is such that it depends only on the commutative subsystem representing the classical controller plus  a quantum noise term.
 In other words, the role of the quantum controller in the feedback loop is equivalent to that of a combination of  the classical controller, the modulator and the measurement devices in the feedback loop.


$\mathbf{Example \  2}$:\ Consider a closed loop system  which
consists of a quantum plant $G$ and a real classical controller $K$
shown in Fig.~\ref{fig:q-c-loop1}. The quantum plant $G$, an optical cavity,  is of the form (\ref{eq:quantum1}) and is given in quadrature form by the equations
\begin{eqnarray}
d q &=& ( -\frac{\gamma}{2} q +\omega p) dt -\sqrt{\gamma} \,  dw_1
\\
dp &=& (-\frac{\gamma}{2} p   -\omega q) dt -\sqrt{\gamma} \,  dw_2
\\
d\eta_1 & = & \sqrt{\gamma}\, q dt + dw_1
\\
d\eta_2 & = & \sqrt{\gamma}\, p dt + dw_2,
\end{eqnarray}
where $\omega$ is the detuning parameter, and $\gamma$ is a coupling constant.
The output of the homodyne detector (Figure \ref{fig:q-c-loop1}) is $\zeta=\eta_1$. The quantum control signal $(w_1,w_2)$  is the output of a modulator corresponding to the equations $d w_1 = \xi dt + d\tilde w_1$, $dw_2 = d \tilde w_2$, 
where $(\tilde w_1, \tilde w_2)$ is a quantum Wiener process, and $\xi$ is a classical state variable  associated with the classical controller $K$, with dynamics $d \xi = - \xi dt + d \zeta$.
The combined hybrid quantum-classical system  $G$-$K$   is given by the equations
\begin{eqnarray}
d q &=& ( -\frac{\gamma}{2} q +\omega p  - \sqrt{\gamma} \, \xi) dt -\sqrt{\gamma} \,  d \tilde w_1
\nonumber \\
dp &=& (-\frac{\gamma}{2} p   -\omega q) dt -\sqrt{\gamma} \,  d \tilde w_2
\nonumber \\
d \xi &=& \sqrt{\gamma} q  dt + d\tilde w_1
\nonumber \\
d\zeta &=&  (\sqrt{\gamma}\, q +  \xi) dt + d\tilde w_1.
\label{eq:eg1-hybrid}
\end{eqnarray}

\begin{figure}[htbp]
 \vspace{-2em}\centering
\includegraphics[scale=0.3]{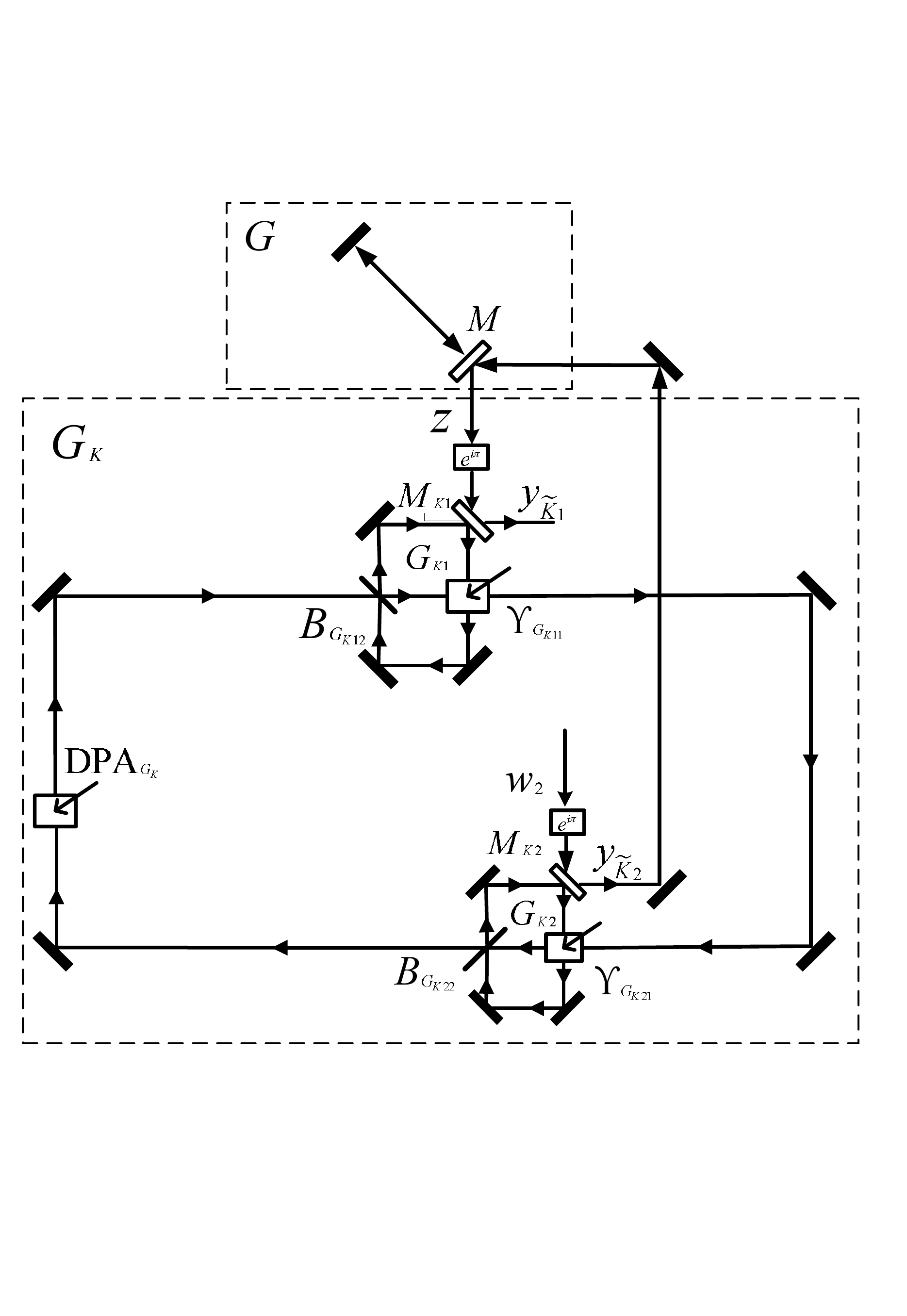}
\vspace{-5em}
\caption{Quantum realization of the closed-loop system shown in Figure \ref{fig:q-c-loop2}.\label{fig:realloop}}
\end{figure}

Note that this hybrid system is an open system, and consequently the equations are driven by quantum noise.
The quantum realization of the system $d \xi = - \xi dt + d \zeta $, $d w_1 =  \xi dt + d\tilde w_1$,
denoted here by $K_Q$ is, from Example 1, given by equations (\ref{eq:eg1}) (with the appropriate notational correspondences).
The combined quantum plant and quantum controller system $G$-$K_Q$  is specified by Figure \ref{fig:q-c-loop2}, with corresponding closed loop equations
\begin{eqnarray}
d q &=& ( -\frac{\gamma}{2} q +\omega p - \sqrt{\gamma}\, x_1) dt -\sqrt{\gamma} \,  dv_2
\nonumber \\
dp &=& (-\frac{\gamma}{2} p   -\omega q  ) dt -\sqrt{\gamma} \,  du_2
\nonumber \\
dx_1 &=& \sqrt{\gamma}\, q  dt + dv_2
\nonumber \\
dx_2 &=& (-x_2 +2 \sqrt{\gamma}\, p )dt  +  du_2 .
\end{eqnarray}
The hybrid dynamics (\ref{eq:eg1-hybrid}) can be seen in these equations (with $x_1$, $v_2$ and $u_2$  replacing $\xi$,  $\tilde w_1$ and $\tilde w_2$, respectively).
By the structure of the equations, joint expectations involving variables in the hybrid quantum plant-classical controller system equal the corresponding expectations for the   combined quantum plant and quantum controller. For example,
$E [ q(t) \xi(t) ] =E[ q(t) x_1(t) ]$. A physical implementation of the new closed loop quantum feedback system is shown in Figure \ref{fig:realloop}.

We consider now the conditional dynamics for the cavity, \cite{WM09,BHJ07}. Let $\hat q(t)$ and $\hat p(t)$ denote the conditional expectations of $q(t)$ and $p(t)$ given the classical quantities  $\zeta(s), \xi(s), \ 0 \leq s \leq t$.
Then
\begin{eqnarray}
d \hat q &=&  ( -\frac{\gamma}{2} \hat q +\omega \hat p -\sqrt{\gamma}   \xi) dt + K_q d\nu
\\
d\hat p &=& (-\frac{\gamma}{2} \hat p   -\omega \hat q) dt  + K_p d\nu��
\end{eqnarray}
where $K_q=\widehat{q^2}-(\hat q)^2 +1$ and $K_p=\widehat{qp}-\hat q \hat p$ are the Kalman gains for the two quadratures, and $\nu$ is the measurement noise (the innovations process, itself a Wiener process).
The output also has the  representation
$d\zeta = (\sqrt{\gamma}\, \hat q +  \xi) dt + d\nu.$
The conditional cavity dynamics combined with the classical controller dynamics leads to the feedback equations
\begin{eqnarray}
d \hat q &=&  ( -\frac{\gamma}{2} \hat q +\omega \hat p -\sqrt{\gamma}  \xi) dt + K_q d\nu
\\
d\hat p &=& (-\frac{\gamma}{2} \hat p   -\omega \hat q) dt  + K_p d\nu\\
d \xi &=&  \sqrt{\gamma} \hat q  dt + d\nu
\\
d\zeta &=&  (\sqrt{\gamma}\, \hat q +  \xi) dt + d\nu
\end{eqnarray}
Here we can see the measurement noise $\nu(t)$ explicitly in the feedback equations.
By properties of conditional expectation, we can relate expectations involving the conditional closed loop system with the hybrid quantum plant classical controller system, e.g.
$E[ \hat q(t) \xi(t) ] = E [ q(t) \xi(t) ]$. We therefore see that the expectations involving the hybrid system, the conditional system, and the quantum plant quantum controller system are all consistent.

\section{Conclusion}
\label{sec:conclusion}
In this paper, we have shown that a class of classical linear
stochastic systems (having a certain form and satisfying certain
technical assumptions) can be realized by quantum linear stochastic
systems.  It is anticipated  that the main results of the work will aid in
facilitation the implementation of classical linear systems with
fast quantum optical devices (eg. measurement feedback control), especially in miniature platforms such as 
nanophotonic circuits.


\addtolength{\textheight}{-3cm}
\end{document}